\documentclass[runningheads]{llncs}
\usepackage[T1]{fontenc}
\usepackage{graphicx}
\usepackage{multirow}
\usepackage{amsfonts}
\usepackage{amsmath}
\usepackage{booktabs}
\usepackage{algorithm}
\usepackage{algorithmic}
\usepackage{makecell}
\usepackage{amsbsy}
\usepackage{multicol}
\usepackage{xcolor,colortbl}
\usepackage{kotex}
\usepackage[colorlinks=true,
            linkcolor=blue,
            urlcolor=blue,
            citecolor=blue]{hyperref}

\begin{document}
%

\title{OCL: Ordinal Contrastive Learning for Imputating Features with Progressive Labels}




%

\author{Seunghun Baek\inst{1}$^*$ 
\and
Jaeyoon Sim\inst{1}$^*$
\and
Guorong Wu\inst{2}
\and
Won Hwa Kim\inst{1}} 
%

\authorrunning{S. Baek et al.}
\titlerunning{OCL: Ordinal Contrastive Learning}


\institute{
Pohang University of Science and Technology, Pohang, South Korea \\
\email{\{habaek4, simjy98, wonhwa\}@postech.ac.kr} \and
University of North Carolina at Chapel Hill, Chapel Hill, USA 
}


%
\maketitle              
\def\thefootnote{*}\footnotetext{Seunghun Baek and Jaeyoon Sim contributed equally to this paper.}



\begin{abstract}

Accurately discriminating progressive stages of Alzheimer's Disease (AD) is crucial for early diagnosis and prevention. 
%
It often involves 
multiple imaging modalities to understand the complex pathology of AD, 
%
however, acquiring a complete set of images is challenging due to high cost and burden for subjects. 
In the end, missing data become inevitable 
which lead to limited sample-size and decrease in precision in downstream analyses. 
%
%
To tackle this challenge, we introduce a holistic imaging feature imputation method 
that
enables to leverage
diverse imaging features while retaining all subjects.
%
The proposed method comprises two networks: 1) An encoder to extract modality-independent embeddings and 2) A decoder to reconstruct the original measures conditioned on their imaging modalities.
%
The encoder includes a novel {\em ordinal contrastive loss}, which aligns samples in the embedding space according to the progression of AD.
%
We also maximize modality-wise coherence of embeddings within each subject, in conjunction with domain adversarial training algorithms, to further enhance alignment between different imaging modalities.
%
The proposed method promotes our holistic imaging feature imputation across various modalities in the shared embedding space. 
%
In the experiments, we show that our networks deliver favorable results for statistical analysis and classification against imputation baselines with 
Alzheimer's Disease Neuroimaging Initiative (ADNI) study. 
\end{abstract}

\section{Introduction}
\label{sec:introduction}


Understanding the progression of neurodegenerative disease such as Alzheimer's Disease (AD) 
often requires a complicated integration of multiple neuroimaging modalities. 
Advanced imaging techniques such as magnetic resonance imaging (MRI) and positron emission tomography (PET) 
provide insights into the structural, functional and molecular changes in the brain associated with AD. 
However, in many clinical studies, it is difficult to collect a complete image set of various imaging modalities
due to cost and participant burden. 
In the end, subjects with missing scans become inevitable that cannot be used for 
downstream analyses, e.g., training a disease prediction model, and they eventually become discarded leading to limited sample-size.   
To learn robust and effective representation to discriminate underlying patterns from progressive disease stages,  
it becomes imperative to accurately impute
the measures of unobserved modalities. 

Recent studies~\cite{translation} translate imaging scans at different levels (e.g., pixel, voxel) by leveraging their correlations~\cite{correlation1,correlation2,correlation3}.
However, focusing on brain regions of interest (ROIs) offers more relevant and interpretable information 
as disease-specific symptoms manifest in arbitrary shapes.  
Region-based models enable the extraction of localized features 
indicative of AD pathology, 
such as hippocampal atrophy or cortical thinning \cite{roi_based1,roi_based2,baek,lsap}. 
Prioritizing these regions enhances the sensitivity of prediction models by effectively capturing subtle disease-specific changes.
Therefore, leveraging region-based models for estimating missing scans holds promise for improving the robustness in AD classification. 

When training a diagnostic model, employing 
Cross-entropy or Supervised Contrastive Learning (SCL) \cite{scl} often 
simply groups the
samples by their labels in an embedding space. 
However, as each AD stage represents the ordinal disease progression such as 
Control (CN), Early Mild Cognitive Impairment (EMCI), Late MCI (LMCI) and AD, 
aligning the 
embedding along the progression continuum 
offers two key advantages. 
First, 
it comprehensively captures the underlying pathology from preclinical stages to full-blown AD,
enriching the data representation 
rather than treating them entirely separate entities. 
Second, it enhances the model robustness to inter-individual variability
in Alzheimer's research 
by focusing on consistent pathological changes rather than individual differences in symptom manifestation. 
To achieve this, we propose {\em ordinal contrastive learning} (OCL), 
which dynamically adjusts the degree of repulsion and attraction among embeddings based on their severity discrepancy informed by disease labels.
Simultaneously,
different modalities from the same individual are pulled together 
to ensure personalized embeddings in the modality-agnostic embedding space. 

With the OCL, we design an architecture 
that projects each sample to a modality-agnostic but progress-aligned embedding space  
and then generates a target measure from the embedding conditioned on a specific imaging modality. 
Our goal is to holistically generate 
realistic data for all missing 
scans personalized to each subject, 
accurately reflecting modality and disease characteristics 
without exhaustively training one-to-one mapping between 
modalities.
\textbf{Key Contributions:} 
{\bf 1)} Our method accurately estimates unobserved imaging measures for individual subjects using their existing data to solidify downstream analyses.
{\bf 2)} We introduce {\em ordinal contrastive learning}, 
which aligns samples in the embedding space based on their disease severity.
{\bf 3)} The experiments on Alzheimer's Disease Neuroimaging Initiative (ADNI) data 
show that our method accurately translates data, capturing realistic information for subsequent analyses.

\section{Method}
\label{sec:method}

\begin{figure}[t!]
\centering
\includegraphics[width=0.96\textwidth]{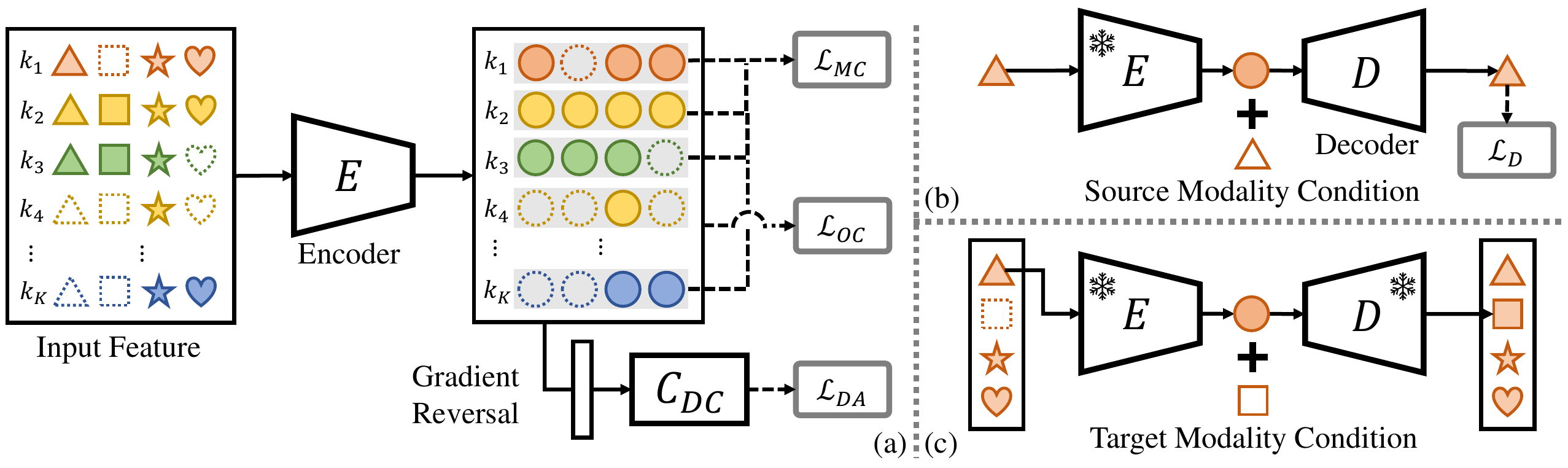}\\
    \caption{\footnotesize 
    Illustration of our framework.
    (a) 
    An encoder $E$ is trained to extract disease progression information across various input modalities through $\mathcal{L}_{DA}$ and $\mathcal{L}_{OC}$. 
    Additionally, $E$ is guided to maximize the similarity of embeddings from the same subject using $\mathcal{L}_{MC}$.
    (b) 
    A decoder $D$ is trained to reconstruct the embedding of a fixed 
    $E$ to its original input under its original modality condition, utilizing 
    $\mathcal{L}_{D}$.
    (c) 
    The trained $E$ and $D$ facilitate the translation of an input to the target modality when the corresponding condition is provided, 
    while preserving disease progression information.
    }
    \label{fig:model_overview}
\end{figure}

In this section, we describe an unified imaging measure-to-measure translation framework for samples with  missing image features. 
Fig.~\ref{fig:model_overview} illustrates the overview of our framework, which consists of two training phases and one inference phase:
(a) An encoder $E$ is trained to map data into a modality-agnostic and disease-progression aligned embedding.
(b) A decoder $D$, conditioned on the source (input) modality, is trained to reconstruct the original measure from the given embedding from (a).
(c) Missing features are imputed from any existing feature of the subject by $E$ and $D$, conditioned on the target (missing) modality.


\subsection{Modality-agnostic Progressive Embedding} 
\label{ssec:training_encoder} 
Let us consider data $X$ consisting of $K$ subjects, 
each with $Q$-dimensional measurements (e.g., $Q$ ROIs) from $S$ imaging features.  
The $k$-th subject $x_k\in\mathbb{R}^{S\times Q}$ has a diagnostic label $y_k \in\{1,\cdots,V\}$ ordered by severity (e.g., CN, EMCI, LMCI and AD), 
and each modality measurement is denoted as $x_{k,s}\in\mathbb{R}^{Q}$. 
For every $k$, 
the encoder $E$ extracts disease progression information informed by $y_k$ from $x_{k,s}$, 
regardless of the modality type $s$. 
This is accomplished by integrating three different guidance strategies as outlined below.

\noindent\textbf{1) Domain Adversarial Training ($\mathcal{L}_{DA}$).} 
We employ a 
domain adversarial training strategy, 
which derives an embedding $z_{k,s}=E(x_{k,s})$ 
with 
$E$ to eliminate modality-specific information associated with $s$. 
This is achieved by a modality classifier $C_{DC}$ and a gradient reversal layer~\cite{domain_adversarial} that reverses the sign of the gradient between $E$ and $C_{DC}$. 
The modality adversarial loss $\mathcal{L}_{DA}$ is defined as
\begin{equation}
\footnotesize
\begin{aligned}
\mathcal{L}_{DA} = \mathcal{J}(s, C_{DC}(E(x_{k,s}))),
\end{aligned}
\end{equation}
where $\mathcal{J}$ represents a suitable loss function (e.g., Cross-entropy). 
Through the gradient reversal layer, 
$E$ is trained to maximize $\mathcal{L}_{DA}$, 
which leads to the discarding of modality-specific information, 
while $C_{DC}$ learns to minimize $\mathcal{L}_{DA}$.

\begin{figure}[t!]
\centering
    \includegraphics[width=0.98\textwidth]{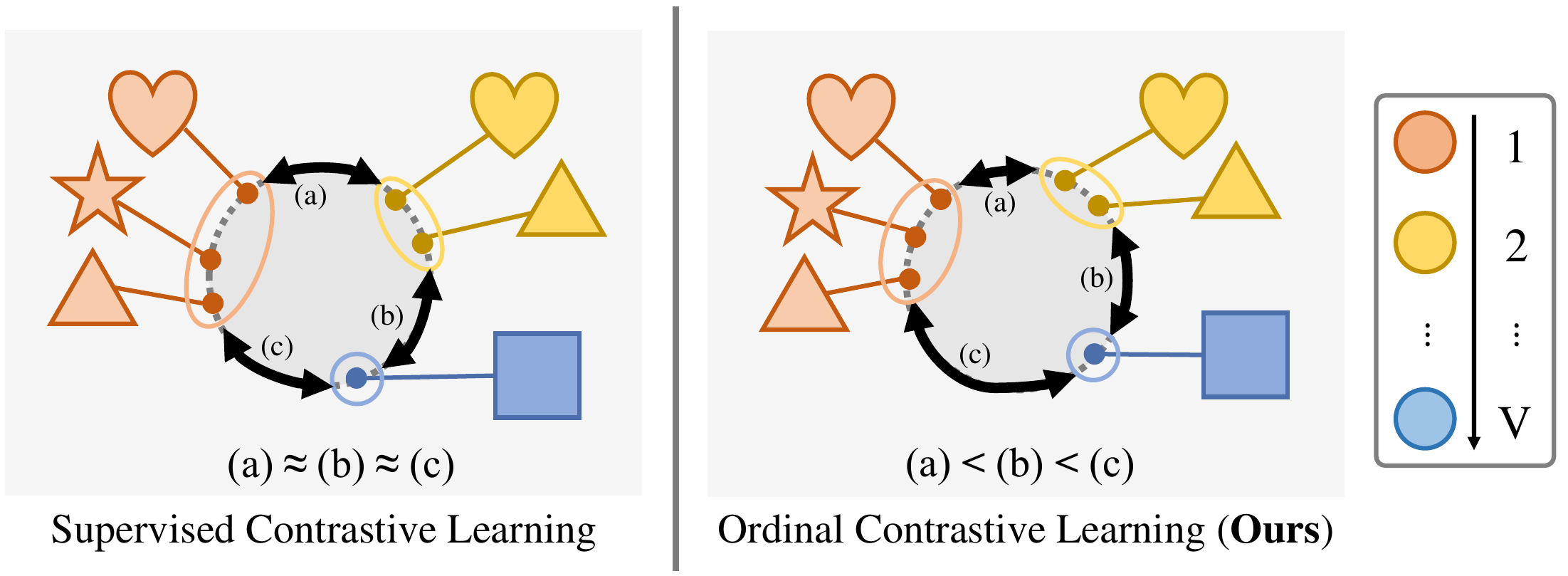}
    \caption{\footnotesize
    Comparison of supervised (left) and ordinal (right) contrastive learning:
    Both approaches contrast the set of all samples from the same class as positives against the negatives from the rest of the batch.
    While supervised contrastive learning repels each negative without differentiation on labels denoted as $(a)\approx(b)\approx(c)$,
    ordinal contrastive learning assigns the penalizing strength based on the label distance.
    }
    \label{fig:ordinal_concept}
\end{figure}

\noindent\textbf{2) Ordinal Contrastive Learning ($\mathcal{L}_{OC}$).}
To accurately characterize measures of missing scans for downstream analyses, 
the encoder $E$ should effectively extract disease progression information. 
$E$ is designed to arrange each sample in the embedding space 
by the orders 
to accurately characterize disease progression.

The OCL is initiated from supervised contrastive learning (SCL)~\cite{scl} which is briefly described below. 
For brevity, we omit the modality type $s\in\{1,\cdots,S\}$ (i.e., $x_{k,\cdot}$ denotes the $k$-th subject) as $\mathcal{L}_{DA}$ removes $s$-specific information in the embedding.
In the context of SCL, within a batch $I$, 
let $i\in I\equiv\{1,...,|I|\}$ represent the index of samples in the batch. 
Then, the embedding $z_{i,\cdot}$ of sample $x_{i,\cdot}$ is supervised by its label $y_i$ through SCL.
When $P(i)\equiv\{p\in P(i):\hat{y_p}=\hat{y}_i\}$ and $N(i)\equiv\{n\in N(i):\hat{y_n}\neq\hat{y}_i\}$ are the sets of indices 
for all positives and negatives in the batch distinct from $i$ each, 
the loss of SCL $\mathcal{L}_{SC}$ is expressed as
\begin{equation}
\footnotesize
\begin{aligned}
    \mathcal{L}_{SC} = \sum_{\substack{i \in I}} \frac{-1}{|P(i)|} \sum_{\substack{p \in P(i)}} \log \frac{\exp(z_i\cdot z_p/\tau)}{\sum\limits_{\substack{p \in P(i)}} \exp(z_i\cdot z_p/\tau)+\sum\limits_{\substack{n \in N(i)}} \exp(z_i\cdot z_n/\tau)}
\end{aligned}
\end{equation}
where $\tau\in\mathbb{R}^+$ is a scalar temperature parameter, 
and $|P(i)|$ is its cardinality.

Notice that a single $\tau$ to control the strength of separation 
acts similar to classification loss, ignoring the degree of differences between each label.
Considering that values of diagnostic label $y\in\{1,\cdots,V\}$ are aligned according to 
their severity (e.g., $i$-th subject is more severe than $n$-th subject if $y_i>y_n$), 
we define a function $d(y_i,y_n)$ measuring the distance between two labels as $|y_i-y_n|$. 
Higher $d(y_i,y_n)$ indicates greater diagnostic differences between $z_i$ and $z_n$. 
Therefore, we make $\tau_{i,n}$ dependent on $y_{i,\cdot}$ and $y_{n,\cdot}$ as $\tau/d(i,n)$ to penalize greater label distance. 
By setting adaptive $\tau_{i,n}$ for each $z_{n,\cdot}$ and unique $\tau_{i,P}$ for every $z_{p,\cdot}$, 
we formulate our ordinal contrastive loss $\mathcal{L}_{OC}$ as
\begin{equation}
\footnotesize
\begin{aligned}
    \mathcal{L}_{OC} = \sum_{\substack{i \in I}} \frac{-1}{|P(i)|} \sum_{\substack{p \in P(i)}} \log \frac{\exp(z_i\cdot z_p/\tau_{i,P})}{\sum\limits_{\substack{q \in P(i)}} \exp(z_i\cdot z_q/\tau_{i,P})+\sum\limits_{\substack{n \in N(i)}} \exp(z_i\cdot z_n/\tau_{i,n})}.
\end{aligned}
\end{equation}
To prevent the collapse or dispersion of the embedding space,
the magnitude of gradient w.r.t positives and negatives should be the same~\cite{how_temperature}.
By the gradient analysis detailed in the supplementary material, 
$\tau_{i,P}$ between $z_i$ and $z_p$ is 
set as 
\begin{equation}
\footnotesize
\begin{aligned}
    \tau_{i,P} = \frac{\sum\limits_{\substack{n \in N(i)}}\exp(z_{i,\cdot}\cdot z_{n,\cdot}\slash\tau_{i,n})}
    {\sum\limits_{\substack{n \in N(i)}} \exp(z_{i,\cdot}\cdot z_{n,\cdot}\slash\tau_{i,n})\slash\tau_{i,n}}.
\end{aligned}
\end{equation}

\noindent\textbf{3) $\mathcal{L}_{MC}$ (Maximize modality-wise coherence within a subject).} 
While $\mathcal{L}_{DA}$ and $\mathcal{L}_{OC}$ 
require only unpaired data, 
identifying pairs originating from the same subject helps alignment
by mapping those pairs to similar embeddings.
To maximize the coherence between modalities of the same subject, 
we design the loss $\mathcal{L}_{MC}$ using
a similarity function $sim(\cdot,\cdot)$ (e.g., cosine similarity) as
\begin{equation}
\footnotesize
\begin{aligned}
    \mathcal{L}_{MC} = \frac{\sum_{k=1}^{K} \sum_{\substack{i,j\in\{1,\cdots,S\} \\ i \neq j}} -\delta_k(i,j) \cdot sim(x_{k,i}, x_{k,j})}{\sum_{k=1}^{K} \sum_{\substack{i,j\in\{1,\cdots,S\} \\ i \neq j}} \delta_k(i,j)}    
\end{aligned}
\end{equation}
where $\delta_{k}(i,j)$ is an indicator function defined as $\delta_{k}(i,j) = 1$ if both $x_{k,i}$ and $x_{k,j}$ exist for subject $k$, and $\delta_{k}(i,j) = 0$ otherwise.
Our final loss to train $E$ is the combination of the three introduced losses defined as
\begin{equation}
\footnotesize
\begin{aligned}
    \mathcal{L}_{E} = \mathcal{L}_{DA} + \mathcal{L}_{OC} + \mathcal{L}_{MC}
\end{aligned}
\end{equation}
where each loss term is equally contributing. 

\subsection{Modality-Conditioned Reconstruction from Embeddings}
\label{ssec:style_injection}

Similar to conditional generation methods
which utilize a single generator to sample different distributions by a conditional vector \cite{cGAN,cVAE}, 
we treat the target modality type $t$ as a one-hot condition vector $c_t\in\mathbb{R}^S$. 
Thus, the decoder $D$ is tasked with estimating values of the target modality from a given embedding under a condition.
Although modality-wise paired data from the same subject are limited, 
due to $\mathcal{L}_{DA}$ and $\mathcal{L}_{MC}$,
the loss for the decoder $\mathcal{L}_{D}$ can be approximated to the self-reconstruction loss of unpaired data $x_{k,t}$ from the translation loss as
\begin{equation}
\footnotesize
    \mathcal{L}_{D}(x_{k,s},x_{k,t}) = ||x_{k,t} - D([E(x_{k,s}),c_t])||^2 \approx
    \mathcal{L}_{D}(x_{k,t}) =  ||x_{k,t} - D([E(x_{k,t}),c_t])||^2 
\end{equation}
where $s,t\in\{1,\cdots,S\}$ correspond to  distinct modality types of source and target each, and [$\cdot$,$\cdot$] denotes to the concatenation of given elements.

\subsection{Imputation Procedure}
\label{ssec:imputation_proedure}

Suppose a subject $k$ lacks a feature $x_{k,t}$.
Our framework 
is capable of generating $\hat{x}_{k,t}$ based on existing $x_{k,s}$ where $s \neq t$, 
leveraging the trained encoder $E$ and decoder $D$ conditioned on the target modality $t$.
This ensures that all $K$ subjects possess feasible features across $S$ modalities to maximize data utilization.

\section{Experiment}
\label{sec:experiment}

\noindent\textbf{Dataset.} 
We validate our framework on four imaging measures from ADNI study~\cite{jack2008alzheimer}. 
Based on Destrieux atlas~\cite{destrieux2010automatic},
each image was partitioned into 148 cortical and 12 sub-cortical regions.
For each parcellation, region-specific imaging features including Standard Uptake Value Ratio (SUVR)~\cite{thie2004understanding} of $\beta$-amyloid protein (AMY), 
metabolism (FDG), and Tau protein (TAU) from PETs,
along with cortical thickness (CT) from MRI scans,
were measured.
The diagnostic labels for each subject include
CN, EMCI, LMCI and AD.
$N$=275 subjects have complete image set denoted as `Common' in Table~\ref{tab:dataset_adni}, 
which serve as the test data in Group Comparison Analysis and Downstream Classification.
To prevent double dipping, 
we exclusively utilize the remaining data to train our framework.

\begin{table*}[!t]
    \caption{\footnotesize Sample-size per modality of ADNI dataset.}
    \centering
    \renewcommand{\arraystretch}{1.0} 
    \renewcommand{\tabcolsep}{0.95cm}
    \scalebox{0.7}{
    \begin{tabular}{c||c|c|c|c||c}
        \Xhline{3\arrayrulewidth}
        \textbf{Label}
        & \multirow{1}{*}{\textbf{CT}} 
        & \multirow{1}{*}{\textbf{TAU}} 
        & \multirow{1}{*}{\textbf{FDG}} 
        & \multirow{1}{*}{\textbf{AMY}} 
        & \multirow{1}{*}{\textbf{Common}} \\ 
        \Xhline{2\arrayrulewidth}
        
        \textbf{CN}    & 844 & 237 & 861 & 735 & 123 \\ \hline
        \textbf{EMCI}  & 490 & 186 & 597 & 833 & 102 \\ \hline
        \textbf{LMCI}  & 250 & 105 & 1138 & 447 & 40 \\ \hline
        \textbf{AD}    & 240 & 85 & 755 & 422 & 10 \\ 
        \hline
        
        \cellcolor{gray!20}\textbf{Total} & 
        \cellcolor{gray!20}1824 & 
        \cellcolor{gray!20}613 & 
        \cellcolor{gray!20}3351 & 
        \cellcolor{gray!20}2437 & 
        \cellcolor{gray!20}275 \\ 
        \Xhline{3\arrayrulewidth}
    \end{tabular}}
    \label{tab:dataset_adni}
\end{table*}

\noindent\textbf{Baselines.}
Our baseline starts from \textit{No Imputation} that utilizes only subjects with all imaging features.
We adopt five established imputation baselines; 
Imputation with diagnostic group means per imaging feature (\textit{Class-wise Mean})~\cite{donders2006gentle},
Linear model-based imputation with chained equations (\textit{MICE})~\cite{mice}, 
Imputation via non-parametric random forests (\textit{MissForest})~\cite{missforest},
Imputation using generative adversarial networks~\cite{GAN} 
(\textit{GAIN})~\cite{gain}, and
Imputation 
with trained models 
utilizing Sinkhorn divergences between random batches for similarity quantification (\textit{Sinkhorn})~\cite{sinkhorn}.
Additionaly, we employ \textit{Pair-wise MLPs},
training one-to-one Multi-Layer Perceptrons (MLPs) for all pairs of 
modalities (i.e., $_{4}P_{2}$ pairs).
For the purpose of ablation, we introduce two additional baselines from our framework (\textit{Ours} ($\mathcal{L}_{OC}+\mathcal{L}_{MC}$)); 
(1) A model that replaces $\mathcal{L}_{OC}$ with $\mathcal{L}_{SC}$ and omits $\mathcal{L}_{MC}$
(\textit{SCL}),
and (2) a model that solely omit $\mathcal{L}_{MC}$ 
(\textit{Ours} ($\mathcal{L}_{OC}$)).


\noindent\textbf{Training.}
For each baselines and our method,
we stack 2 layers with 128 hidden units for $E$ and $D$ each.
AdamW~\cite{AdamW} with learning rate $10^{-3}$ is used, decayed at 0.05 for every layer.
We train each model 3000 epochs with a batch size of 4096.

\section{Result and Analysis}

\begin{figure}[t!]
    \centering
    \scriptsize
    \renewcommand{\arraystretch}{0.8}
    \renewcommand{\tabcolsep}{0.1cm}
        \scalebox{0.96}{
        \begin{tabular}{cccc}
            & \raisebox{0\height}[0pt][0pt]{$\mathcal{L}_{CE}+\mathcal{L}_{DA}$} 
            & \raisebox{0\height}[0pt][0pt]{$\mathcal{L}_{SC}+\mathcal{L}_{DA}$}
            & \raisebox{0\height}[0pt][0pt]{$\mathcal{L}_{OC}+\mathcal{L}_{DA}$ {\bf (Ours)}} \\

            \raisebox{6\height}[0pt][0pt]{(a)} 
            & \includegraphics[width=0.3\linewidth]{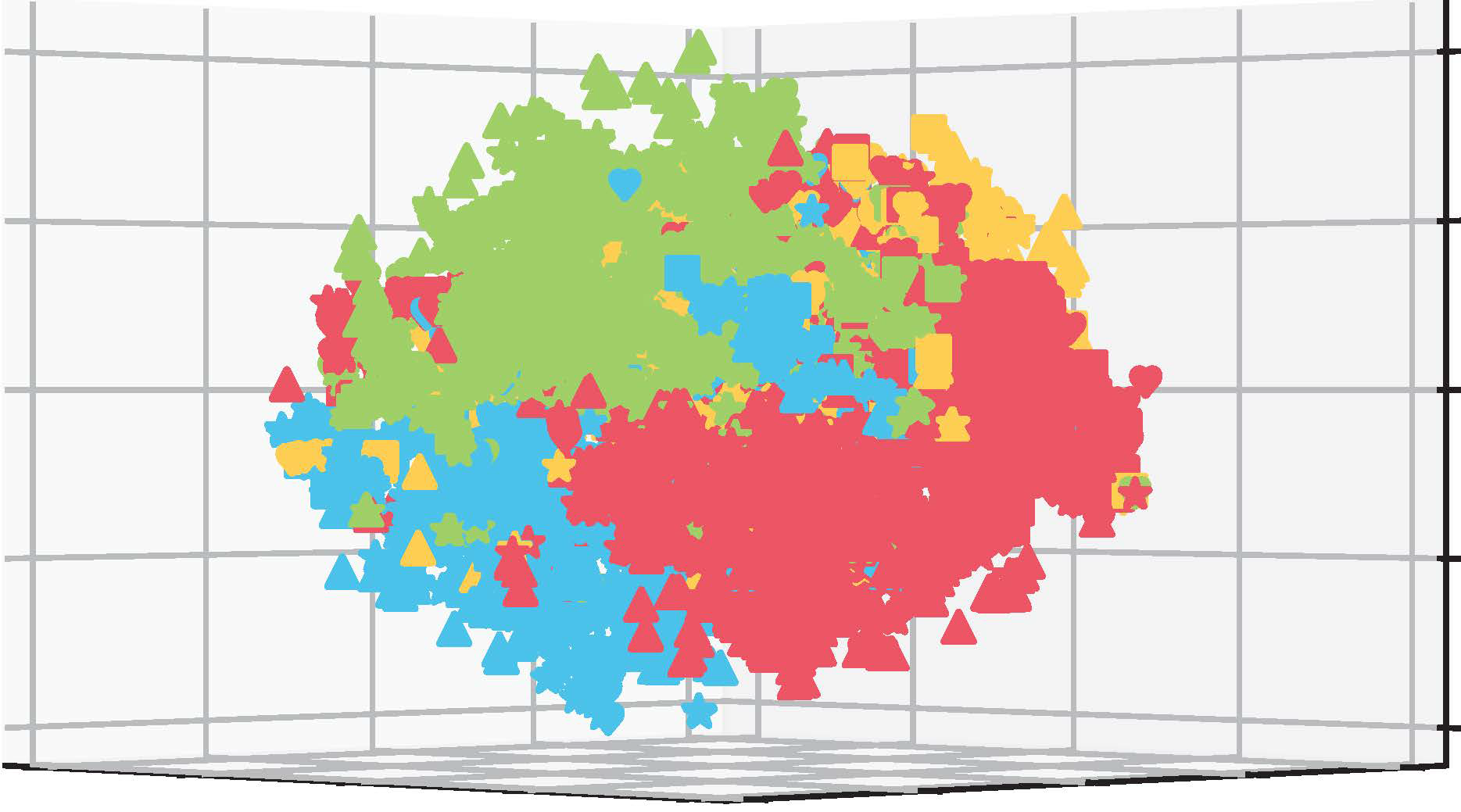}
            & \includegraphics[width=0.3\linewidth]{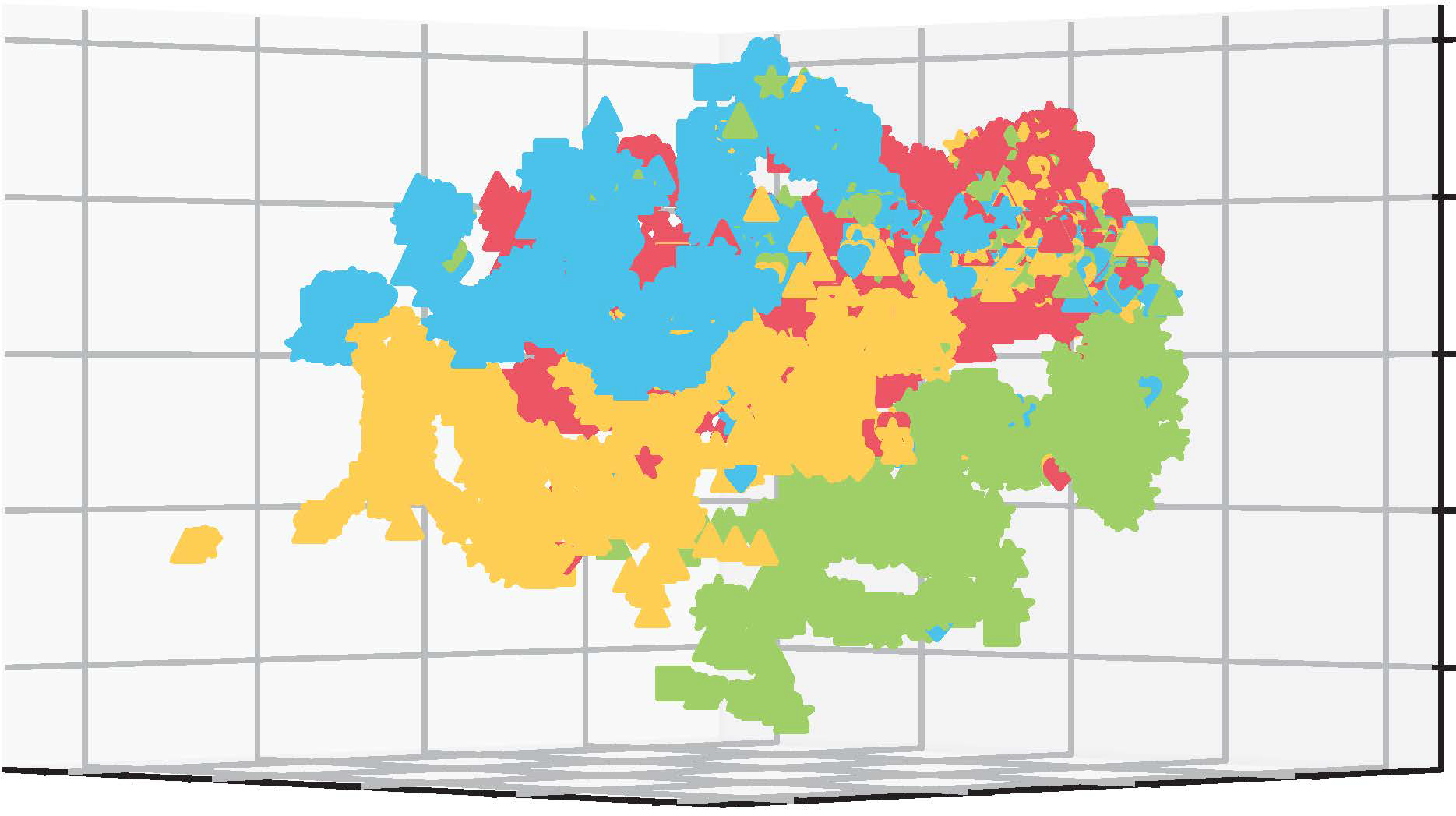}
            & \includegraphics[width=0.3\linewidth]{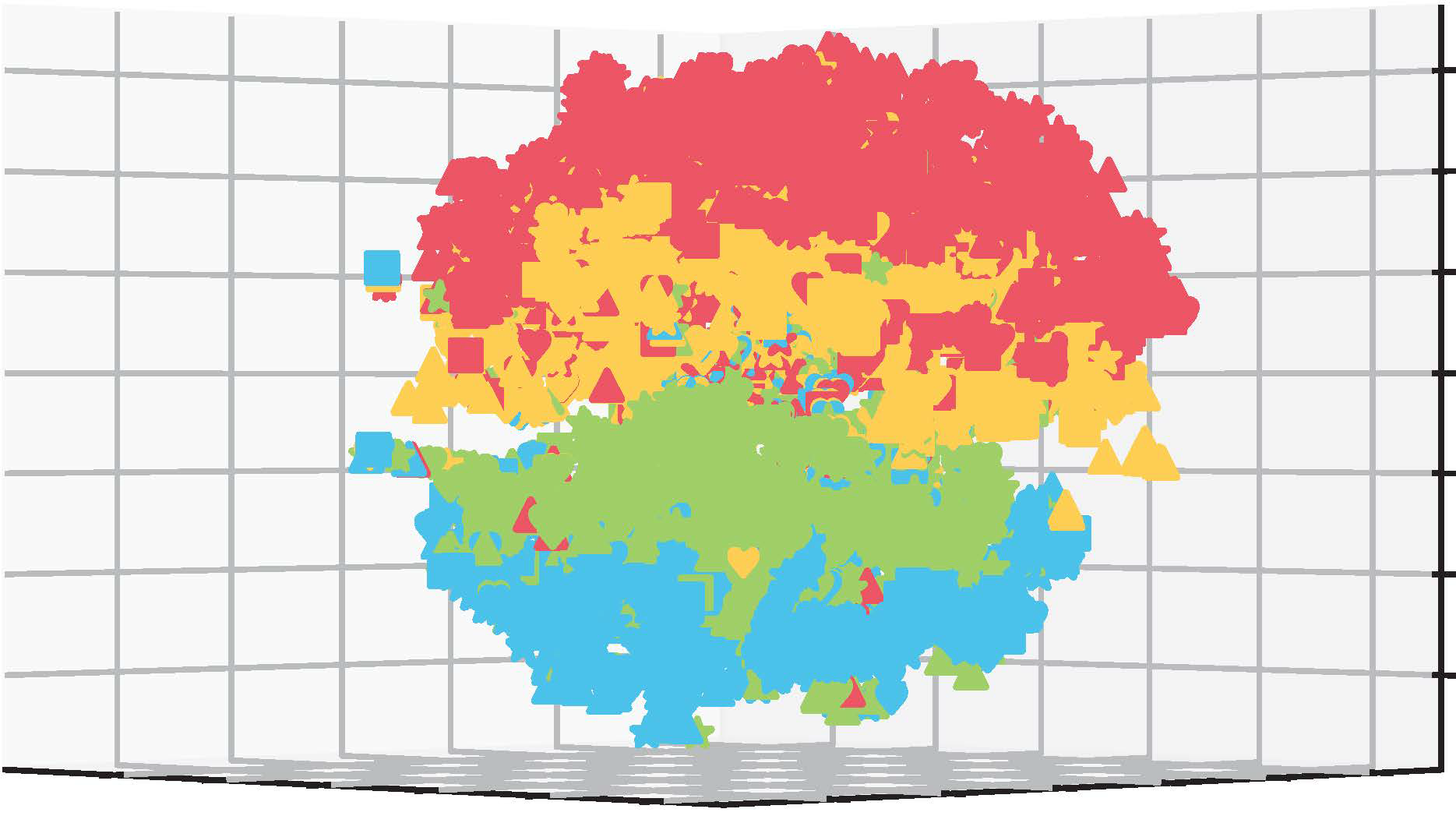}\\

            \raisebox{6\height}[0pt][0pt]{(b)}
            & \includegraphics[width=0.3\linewidth]{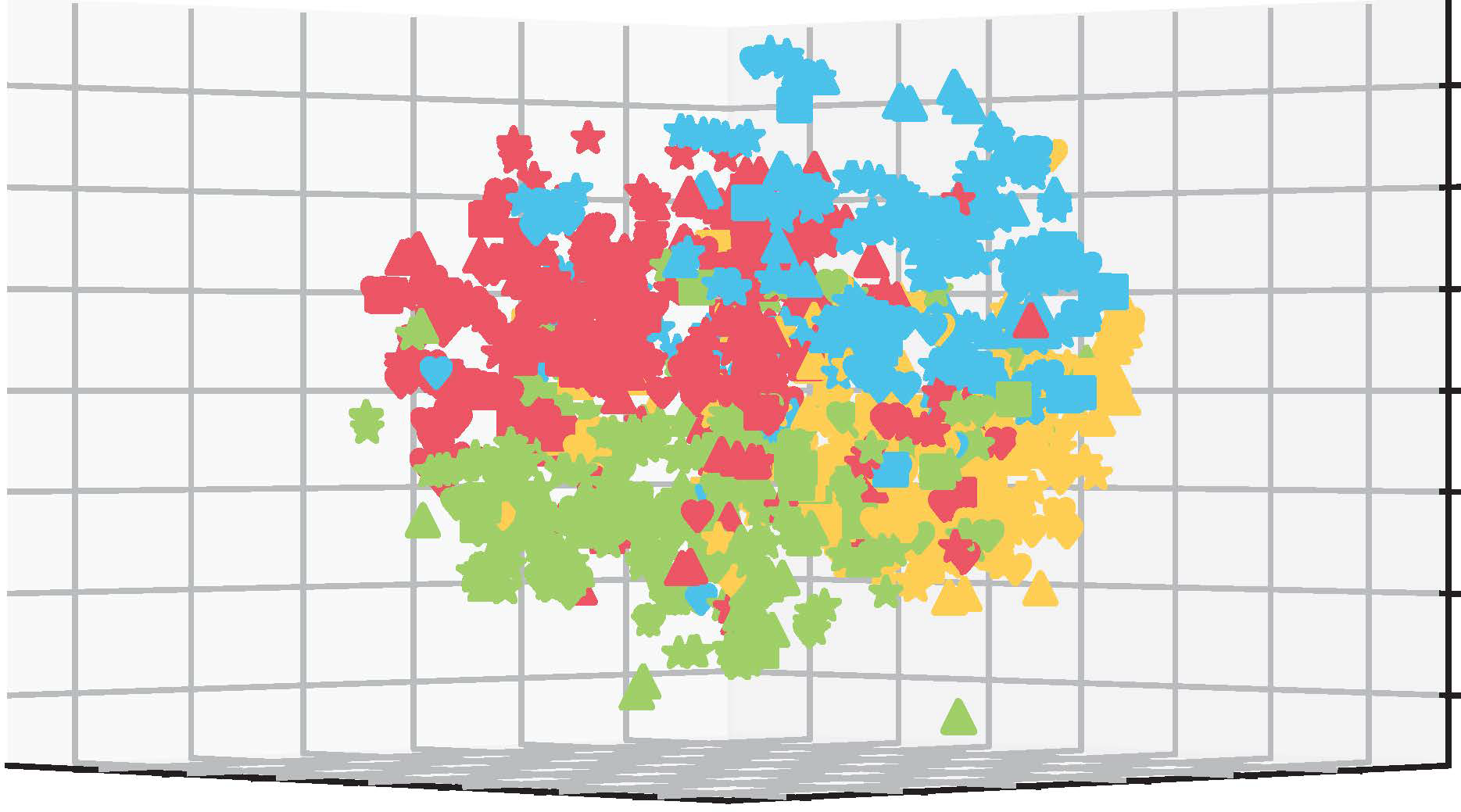}
            & \includegraphics[width=0.3\linewidth]{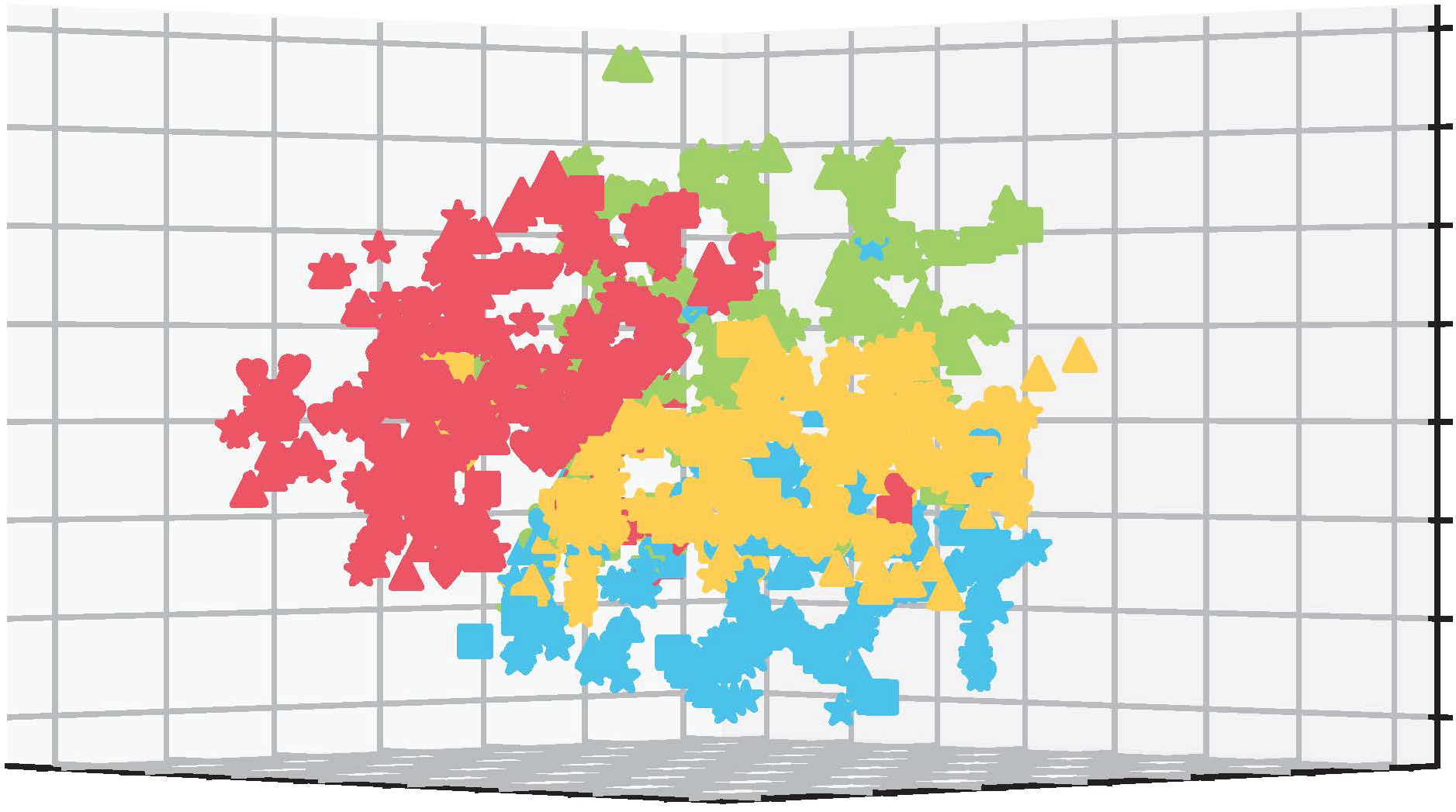}
            & \includegraphics[width=0.3\linewidth]{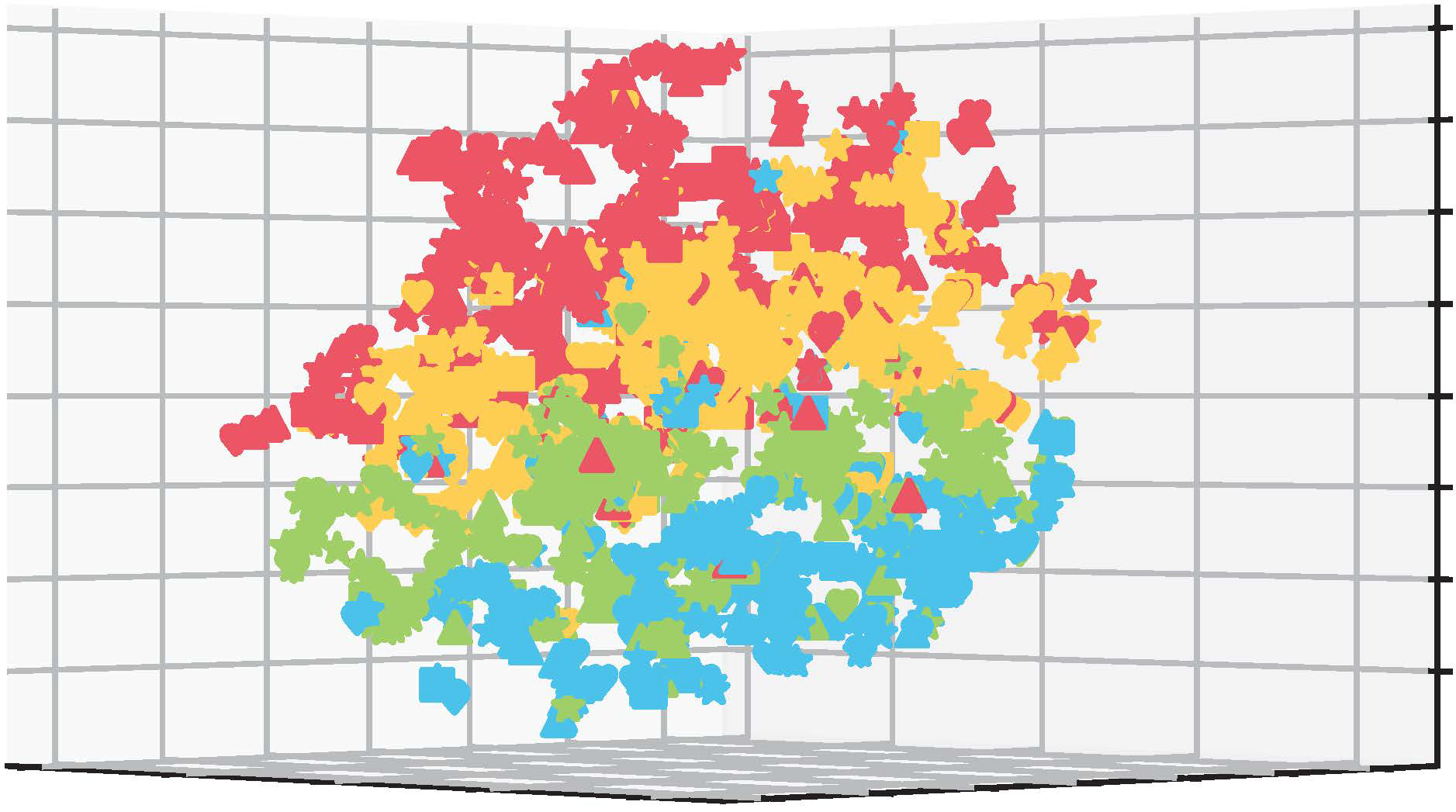}\\
            
            & \multicolumn{3}{c}{\includegraphics[width=0.92\linewidth]{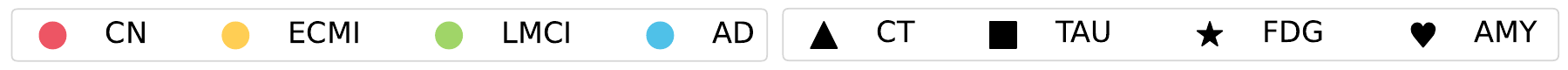}}\\
        \end{tabular}}
    \caption{\footnotesize 
    Visualizations of embeddings under each loss by t-SNE~\cite{tsne}.
    Each individual encoder is trained with three distinct losses including 
    Cross-Entropy $\mathcal{L}_{CE}$ (left), Supervised Contrastive Loss $\mathcal{L}_{SC}$ (center) and our Ordinal Contrastive Loss $\mathcal{L}_{OC}$ (right)
    along with domain adversarial loss $\mathcal{L}_{DA}$.
    (a) and (b) correspond to training and testing data respectively.
    (Color: AD-stage labels, Shape: imaging scan types.)
    }
    \label{fig:embeddings}
\end{figure}

\noindent\textbf{Embedding Space Analysis.}
To verify  $\mathcal{L}_{OC}$ for aligning samples in the embedding space,
we visualize the embeddings using t-SNE~\cite{tsne} 
compared to those from Cross-entropy $\mathcal{L}_{CE}$ and supervised contrastive learning $\mathcal{L}_{SC}$.
As there are only 10 AD subjects in the test set (i.e., `Common' in the Table~\ref{tab:dataset_adni}), 
we randomly split the whole data in 8:2 ratio for train/test sets only in this visualization experiment.
For every result in Fig.~\ref{fig:embeddings}, 
$\mathcal{L}_{DA}$ was adopted together with each loss to utilize single $E$ for every modality.
As $\mathcal{L}_{CE}$ and $\mathcal{L}_{SC}$ simply separate the samples without differentiation
under diagnostic label in the embedding space,
resulting embeddings do not show any visible order.
However, our $\mathcal{L}_{OC}$ aligns the embedding under the disease severity as we argued in section~\ref{ssec:training_encoder}.

\begin{figure*}[!b]
    \centering

    


    \renewcommand{\arraystretch}{0.5}
    \renewcommand{\tabcolsep}{0.01cm}
    \scalebox{1.0}{\scriptsize
    \begin{tabular}{ccccc}
    
     \includegraphics[width=0.225\linewidth]{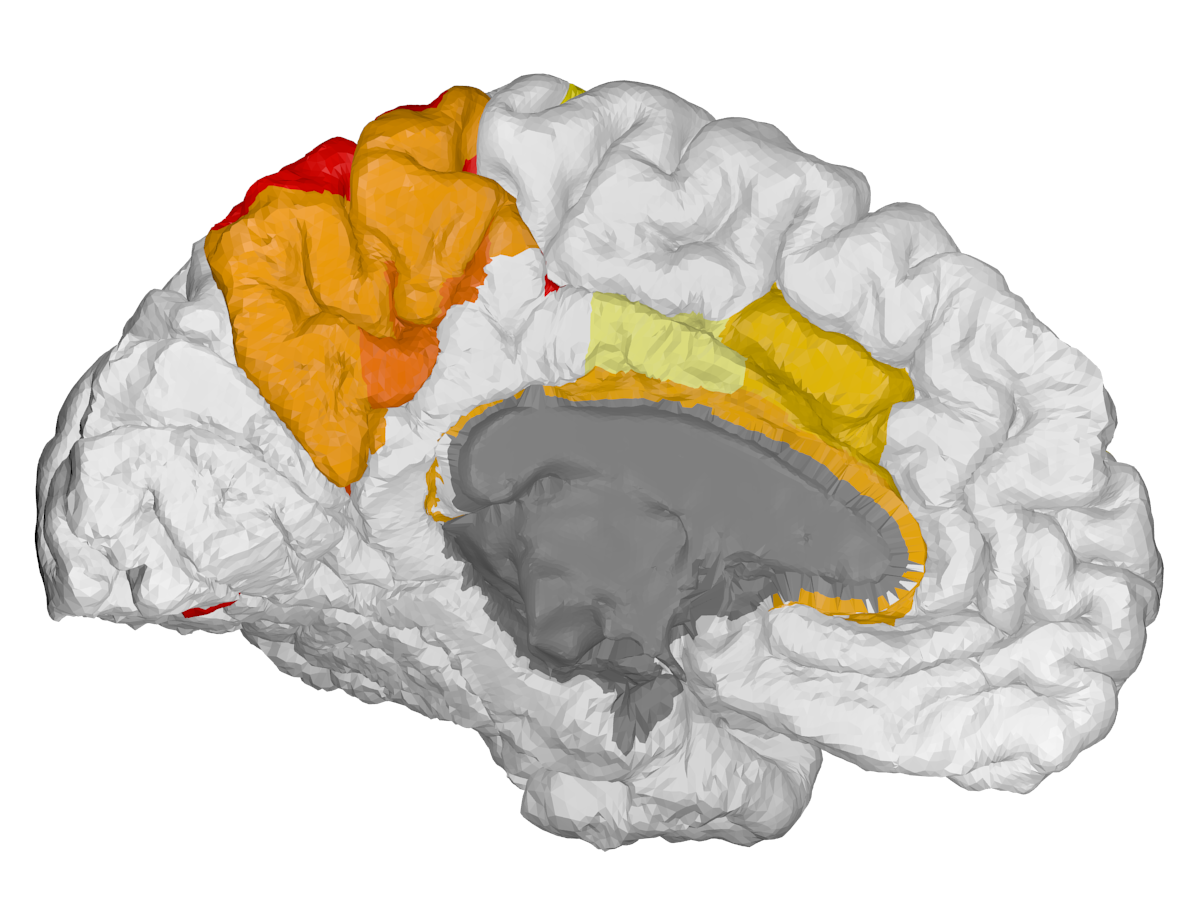}
    &\multicolumn{1}{c@{\hspace{0.075cm}}|}{\includegraphics[width=0.225\linewidth]{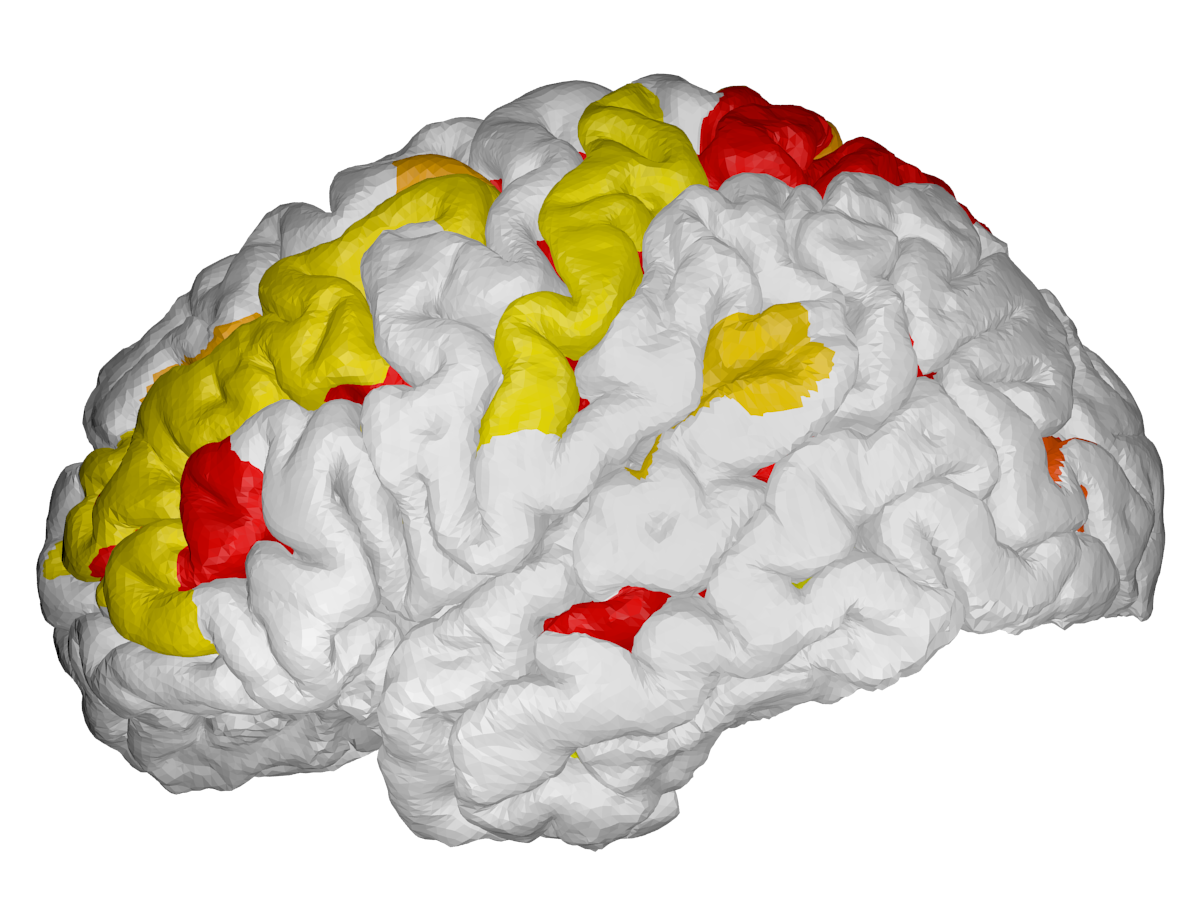}}
    &\multicolumn{1}{@{\hspace{0.075cm}}c}{\includegraphics[width=0.225\linewidth]{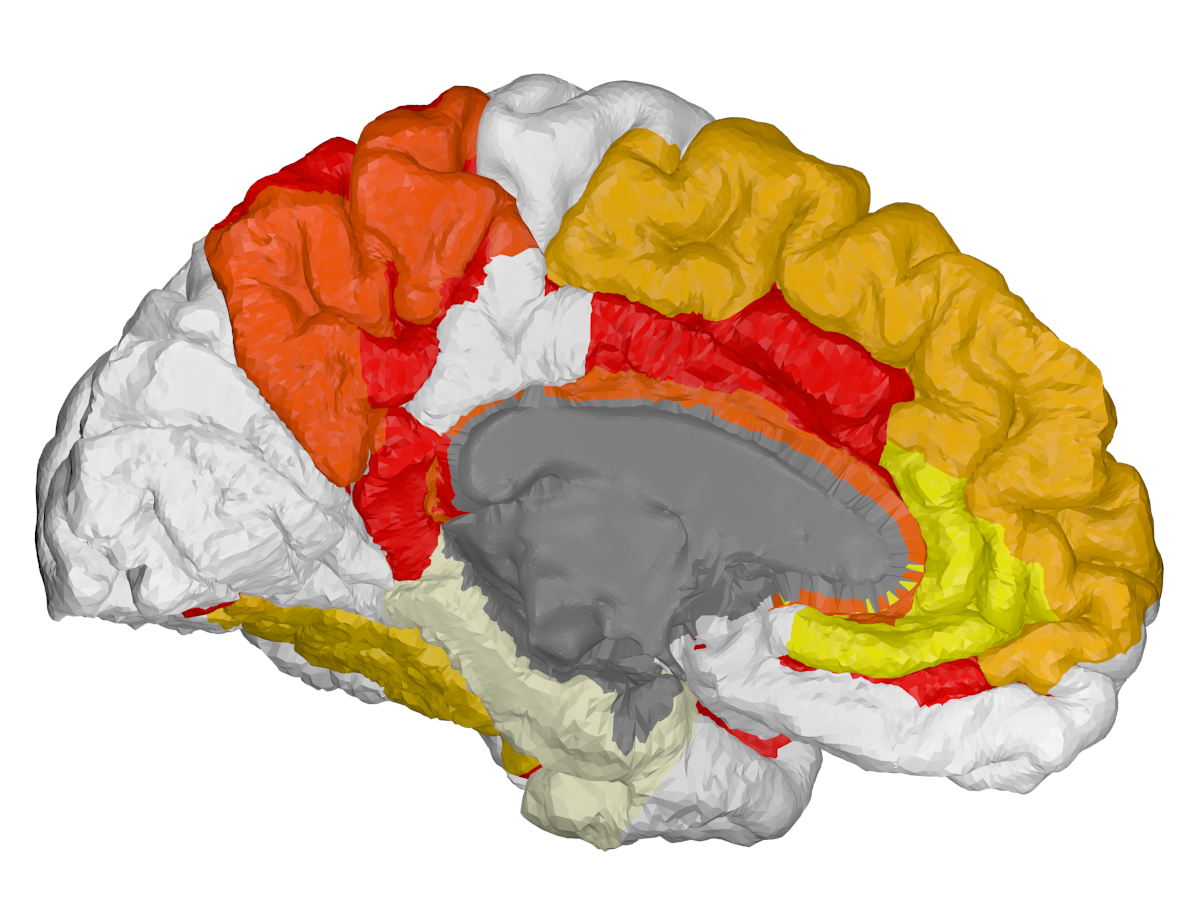}}
    &\includegraphics[width=0.225\linewidth]{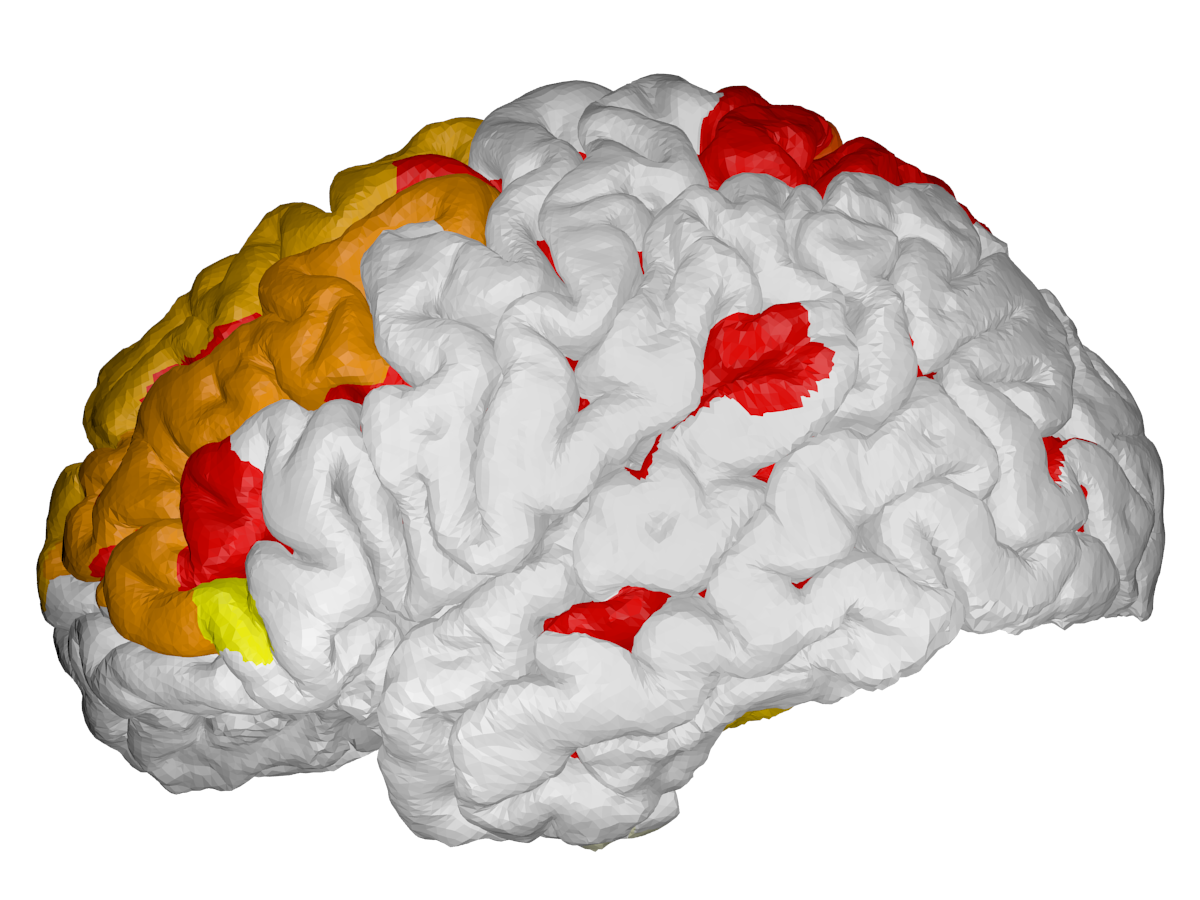}
    &\\

    \multicolumn{2}{c|}{\quad\raisebox{0\height}[0pt][0pt]{(a) Before imputation}}
    & \multicolumn{2}{c}{\quad\raisebox{0\height}[0pt][0pt]{(b) After imputation}}
    &\quad\raisebox{0\height}[0pt][0pt]{\includegraphics[width=0.043\linewidth]{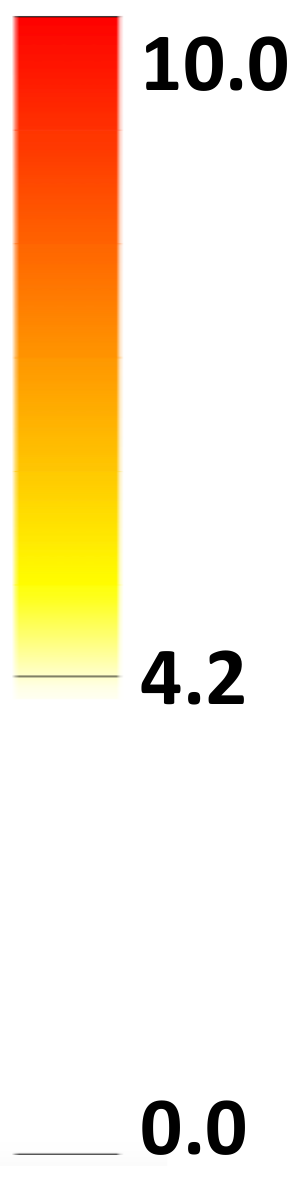}}\\
    \\
    \end{tabular}}
    
    \renewcommand{\arraystretch}{1.0}
    \renewcommand{\tabcolsep}{0.65cm}
    \scalebox{0.7}{
    \begin{tabular}{l||c|c|c|c|c|c}
        \Xhline{3\arrayrulewidth}
        \multirow{2}{*}{\textbf{Modality}}
        & \multicolumn{2}{c|}{\bf CN vs EMCI}
        & \multicolumn{2}{c|}{\bf EMCI vs LMCI}
        & \multicolumn{2}{c}{\bf LMCI vs AD}\\
        \cline{2-7}
        
        & {\bf (a)} & {\bf (b)} 
        & {\bf (a)} & {\bf (b)} 
        & {\bf (a)} & {\bf (b)} \\
        \Xhline{2.5\arrayrulewidth}
        
        \textbf{Cortical Thickness} 
        & 59 & 88 (57)
        & 24 & 64 (20)
        & 55 & 131 (55)\\
        \hline

        \textbf{Tau} 
        & 0 & 84 (0)
        & 1 & 22 (1)
        & 9 & 99 (9)\\
        \hline

        \textbf{FDG}
        & 48 & 83 (44)
        & 77 & 94 (75)
        & 139 & 119 (119)\\
        \hline

        \textbf{$\beta$-Amyloid}
        & 32 & 70 (27)
        &  6 & 78 (6)
        & 144 & 152 (144)\\
        \Xhline{3\arrayrulewidth}
    \end{tabular}}
\caption{\footnotesize
$p$-values from group comparisons with Bonferroni correction at $\alpha=0.01$: 
(a) before imputation, (b) after imputation from our model. 
Top: Resutant $p$-value maps on a brain surface (left hemisphere)  \cite{brainpainter} in a $-log_{10}$ from CN and EMCI comparison with cortical thickness, and (b) shows higher sensitivity. 
Bottom: Number of significant ROIs. Number of common ROIs before-and-after imputation are in ().
}
\label{fig:t_test}
\end{figure*}

\noindent\textbf{Statistical Analysis.}
To validate the robustness of our estimated data, 
we conduct statistical group comparisons at each ROI-level between consecutive groups (i.e., CN-EMCI, EMCI-LMCI, LMCI-AD). 
Our goal is to detect more statistically significant ROIs by imputing the missing features with our method.
The increase indicates higher statistical sensitivity than using only real samples, 
as randomly generated data will not change the effect size between the groups. 

In Fig. \ref{fig:t_test}, we summarize the count of significant ROIs, whose $p$-values survive multiple comparison correction i.e., Bonferroni correction at 0.01 \cite{bonferroni}. 
With the imputation, 
the number of significant ROIs increases in most experiments depicted in Fig. \ref{fig:t_test} (bottom). 
The Fig. \ref{fig:t_test} (top) demonstrates 
the results of CN vs. EMCI analysis with CT on a cortical surface, which highlights the enhanced sensitivity (i.e., lower $p$-values) from the imputation in several frontal and temporal regions \cite{sig_ad}.  
Since the resulting ROIs from our method subsume the majority of the ROIs from the real data analysis, 
as summarized in () of Fig. \ref{fig:t_test}, 
it indicates that our approach preserves the characteristics of the real data as well.
Comparing Fig. \ref{fig:t_test} (a) and (b) at the top, the ROIs newly identified includes {\em superior frontal gyrus}, {\em superior temporal gyrus} and {\em anterior cingulate cortex} which are 
major ROIs in preclinical AD analysis \cite{superior_frontal,superior_temporal,anterior_cingulate}.

\begin{table*}[t!]
\caption{\footnotesize Classification performance 
on ADNI data with all imaging features. 
}
\centering
\renewcommand{\arraystretch}{1.0}
\renewcommand{\tabcolsep}{0.18cm}
\scalebox{0.7}{
\begin{tabular}{l||ccc|ccc}
    \Xhline{3\arrayrulewidth}
    \textbf{Classifier} & \multicolumn{3}{c|}{\textbf{MLP (2 layers)}} & \multicolumn{3}{c}{\textbf{MLP (4 layers)}} \\
    \hline

    \textbf{Method} & \textbf{Accuracy} & \textbf{Precision} &  \textbf{Recall} & \textbf{Accuracy} & \textbf{Precision} &  \textbf{Recall} \\
    \Xhline{2.5\arrayrulewidth}
    
    
    No Imputation
    & 0.673$\pm$0.030 & 0.659$\pm$0.025 & 0.673$\pm$0.030 
    & 0.698$\pm$0.048 & 0.707$\pm$0.047 & 0.698$\pm$0.048 \\
    \hline
    
    Class-wise Mean~\cite{donders2006gentle}
    & 0.753$\pm$0.050 & 0.778$\pm$0.041 & 0.753$\pm$0.050 
    & 0.775$\pm$0.036 & 0.771$\pm$0.032 & 0.775$\pm$0.036 \\

    MICE~\cite{mice}
    & 0.739$\pm$0.043 & 0.761$\pm$0.046 & 0.739$\pm$0.043 
    & 0.814$\pm$0.043 & 0.761$\pm$0.046 & 0.814$\pm$0.043 \\

    MissForest~\cite{missforest}
    & 0.721$\pm$0.061 & 0.753$\pm$0.080 & 0.721$\pm$0.061 
    & 0.832$\pm$0.025 & 0.844$\pm$0.024 & 0.832$\pm$0.025 \\

    Sinkhorn~\cite{sinkhorn}
    & 0.776$\pm$0.044 & 0.799$\pm$0.041 & 0.776$\pm$0.044 
    & 0.829$\pm$0.033 & 0.847$\pm$0.041 & 0.829$\pm$0.033 \\
    
    GAIN~\cite{gain}
    & 0.752$\pm$0.029 & 0.766$\pm$0.022 & 0.752$\pm$0.029 
    & 0.795$\pm$0.054 & 0.805$\pm$0.050 & 0.795$\pm$0.054 \\
    
    Pair-wise MLPs
    & 0.756$\pm$0.030 & 0.782$\pm$0.036 & 0.756$\pm$0.030 
    & 0.782$\pm$0.062 & 0.799$\pm$0.060 & 0.782$\pm$0.062 \\

    

    SCL~\cite{scl}
    & 0.813$\pm$0.042 & 0.812$\pm$0.051 & 0.813$\pm$0.042 
    & 0.845$\pm$0.020 & 0.851$\pm$0.038 & 0.845$\pm$0.020 \\\hline

    \cellcolor{gray!20}Ours 
    ($\mathcal{L}_{OC})$
    & \cellcolor{gray!20}0.826$\pm$0.029 & \cellcolor{gray!20}0.829$\pm$0.021 & \cellcolor{gray!20}0.826$\pm$0.029 
    & \cellcolor{gray!20}0.851$\pm$0.046 & \cellcolor{gray!20}0.862$\pm$0.051 & \cellcolor{gray!20}0.851$\pm$0.046 \\

    \cellcolor{gray!20}Ours ($\mathcal{L}_{OC}+\mathcal{L}_{MC}$)
    & \cellcolor{gray!20}\textbf{0.829$\pm$0.042} 
    & \cellcolor{gray!20}\textbf{0.839$\pm$0.041} 
    & \cellcolor{gray!20}\textbf{0.829$\pm$0.042} 
    & \cellcolor{gray!20}\textbf{0.854$\pm$0.025} 
    & \cellcolor{gray!20}\textbf{0.862$\pm$0.024} 
    & \cellcolor{gray!20}\textbf{0.854$\pm$0.025} \\
    \Xhline{3\arrayrulewidth}

    \end{tabular}}
\label{tab:classification}
\end{table*}


\noindent\textbf{Downstream Classification Performance.}
We employ MLPs with 2 and 4 layers 
as downstream 
classifiers
to demonstrate 
effectiveness of our imputation. 
These models take whole imaging measures as input after concatenation  (e.g., $x_k\in\mathbb{R}^{S\cdot Q}$ for $k$-th subject).
For 
unbiased results, we perform classification using 
cross-validation (CV). 
Beginning with the $No~Imputation$ scenario, 
we partition samples from common subjects into 5 folds, 
with one fold served as test data while the rest folds are utilized to train the classification model.
For other methods, 
subjects with missing features are integrated as supplementary training data for the classification models 
after applying each distinctive imputation method.

In Table~\ref{tab:classification}, 
we report the classification performance derived from the 5-fold CV.
Imputation improved performance over 
\textit{No Imputation} highlighting the advantage of utilizing 
incomplete subjects. 
Most of all, our framework outperformed all imputation baselines across both classifiers, 
showing superior results for all evaluation metrics.
When comparing the use of $\mathcal{L}_{SC}$ (for SCL) and $\mathcal{L}_{OC}$, 
utilizing 
$\mathcal{L}_{OC}$ yields better performance by 1.3\%p in accuracy, 
indicating that $\mathcal{L}_{OC}$ helps imputing data with ordinal labels.
In addition, $\mathcal{L}_{MC}$ also improves the result by personalizing the embeddings from different modalities of the same subject, achieving almost 83\% accuracy in the 4-way classification, which would remain as 67.3\% without the our imputation.    

\section{Conclusion}
\label{sec:conclusion}

In this work, we propose a promising framework that imputes unobserved imaging measures of subjects
by translating their existing measures.
To enable holistic imputation accurately reflecting individual disease conditions,
our framework devises 
modality-invariant and disease-progress aligned latent space 
guided by \textbf{1)} domain adversarial training, \textbf{2)} maximizing modality-wise coherence, and \textbf{3)} ordinal contrastive learning. 
Experimental results on the ADNI study 
show that our model offers reliable estimations of unobserved modalities 
for individual subjects,
facilitating the downstream AD analyses.
Our work has potential to be adopted by other neuroimaging studies 
suffering from missing measures.

%
%

\begin{credits}
\subsubsection{\ackname} 
This research was supported by 
NRF-2022R1A2C2092336 (50\%), 
RS-2022-II2202290 (20\%), 
RS-2019-II191906 (AI Graduate Program at POSTECH, 10\%) funded by MSIT,
RS-2022-KH127855 (10\%), 
RS-2022-KH128705 (10\%) funded by MOHW from South Korea.

\subsubsection{\discintname}
The authors have no competing interests to declare that are
relevant to the content of this article.

\end{credits}

\bibliographystyle{splncs04}
\bibliography{Refs-1615}




\end{document}